# Negative Refraction Does Not Make Perfect Lenses

Weiguo Yang

Department of Engineering & Technology

Western Carolina University

Cullowhee, NC 28723, USA

wyang@wcu.edu

Michael A. Fiddy

Center for Optoelectronics and Optical Communications

University of North Carolina at Charlotte

Charlotte, NC 28223, U.S.A.

mafiddy@uncc.edu

Abstract: The widely-accepted theoretical treatment of the electromagnetic boundary problem of evanescent wave transfer at an interface between a normal medium of n=1 and an ideal negative index medium of n=-1 neglects the non-zero induced surface current and charge densities at the interface and is self-inconsistent. We re-solve the electromagnetic boundary problem by taking into account the non-zero induced surface current and charge densities that have been neglected so far by others. We give the exact induced surface current and charge distributions for this special case and solve the refracted and reflected fields analytically using Green's function method. The self-consistent solution yields a transmission coefficient of 1 and reflection coefficient of 0 for all evanescent waves. Accordingly, we found that, on the contrary to the popular belief, negative index of refraction does not make perfect lenses.

## I. INTRODUCTION

Pendry's seminal paper [1] on the perfect lens made of a slab of negative index medium (NIM) with $\varepsilon$=-1 and $\mu$=-1 has inspired an exciting new research field in negative index materials and, more generally, metamaterials. These unusual materials have shown great promise in many breakthrough applications such as super-resolution imaging and invisibility cloaking [2]–[19]. Metamaterials also have shown increased interests in other practical applications such as electronically small antennas, enhancement of solar cell absorption and LD/LED light extraction, photonic density of state engineering and radiation control [20]–[24]. With the advancement in nano-fabrication technology, the metamaterials research has seen and will certainly continue to see many more fruitful applications in the future. However, the widely-accepted theory of Pendry for a NIM perfect lens neglects the effect of non-zero surface current and charge densities at the NIM interface and is self-inconsistent. Initially, Pendry's surprising result that a passive slab of negative index medium with $\epsilon$=-1 and $\mu$=-1 can amplify all evanescent waves indefinitely was hotly debated. 't Hooft commented that Pendry might have used incorrect arguments to arrive at the otherwise correct conclusions [25]. 't Hooft did also point out a puzzling fact of Pendry's theory that, due to the amplification of the evanescent waves, the amplitude of the electric field can grow extremely large and seems "can easily reach values beyond the breakdown of any material". In reply, Pendry argued that the approach he took in his original paper is "in accord with multiple scattering theory as documented in standard textbooks" [26]. As for the puzzling point of possible infinitely large energy density, Pendry argued that since evanescent waves do not transport energy, therefore, energy conservation is not violated by amplification of evanescent waves. Although there is no public record showing that this conversion occurred, Pendry's reply to the diverging energy density issue is unsatisfactory. This is because, although evanescent waves do not transport energy, they do possess electro-magnetic field energy. With exponentially growing amplitude, the electromagnetic field energy density due to the presence of these evanescent waves does grow exponentially. The fact that evanescent waves ~~does~~ possess electromagnetic energy was also pointed out by Williams [27]. Williams had several other issues with the Pendry's theory but all have been well-addressed in Pendry's reply [28]. Garcia and Nieto-Vesperinas [29]–[31] argued as early as 2002 that although there is amplification of evanescent waves in the ideal lossless, dispersive-less NIM, they still do not make a perfect lens since the effect is limited to a finite thickness of the slab and therefore prevents image forming. Furthermore, Garcia and Nieto-Vesperinas pointed out that any loss may dramatically diminish the evanescent wave amplification effect and instead change it to decay [31]. Pendry himself, in collaboration with Smith and colleagues [32], has also addressed the lossy NIM issue and concluded that the loss may indeed pose a severe limitation to the image resolution as well as the thickness of the NIM slab lens, limiting it to only a small fraction of a wavelength with any practical material loss. Much of these early debates were subdued, especially after the experimental results, notably by Liu, et al. [33], indicating the support for the amplification of evanescent waves in a silver slab as originally suggested in Pendry's paper. Since 2008, we raised another issue with Pendry's perfect lens theory [34]–[36]. The issue centered around our first realization that the usual electromagnetic boundary conditions, namely, the tangential components of both the electric field and magnetic field are continuous across the interface, which are normally applied in all previous discussions regarding the NIM, is self-contradictory when evanescent waves exist at the interface between air and ideal NIM of $\epsilon$ =-1 and $\mu$ =-1. This realization, however, went largely un-noticed. In the manuscript reviewing process that followed as well as private communications with Pendry, the issue of abnormal boundary conditions has been largely ignored, sometimes by citing that the theory has been experimentally verified and there is no need for further discussion. Recently we came to realize that in the special case of evanescent wave refraction at the interface of air and an ideal NIM, exact induced surface current and charge distributions can be retrieved [36]. Accordingly, the transferred electromagnetic field can be solved analytically using the Green's function method knowing the field sources at the boundary. This self-consistent solution yields a transmission coefficient of 1 and reflection coefficient of 0 for all evanescent waves. As a result, the evanescent waves will not be amplified by the NIM slab. Accordingly, we conclude that, on the contrary to popular belief, a negative index of refraction does not make perfect lenses.

In the following sections, we will first demonstrate the contradiction in Pendry's perfect lens theory due to the neglect of possible surface currents and surface charges. Then we will derive the exact distributions of the induced surface current and charge densities and from these source terms, we give the solutions to the refracted waves and hence the transmission and reflection coefficients. It then can be shown that for all evanescent waves the transmission coefficient is 1 and the reflection coefficient is 0, which leads to the conclusion that n = -1 does not make a perfect lens.

## II. SELF-CONTRADICTION IN PENDRY'S PERFECT LENS THEORY

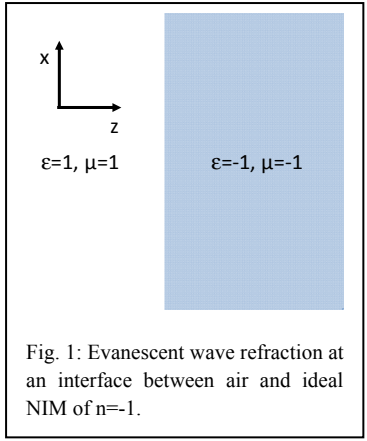

Fig. 1: Evanescent wave refraction at an interface between air and ideal NIM of n=-1.

Consider the case of an interface between air and an ideal NIM with relative permittivity $\epsilon = -1$ and relative permeability $\mu = -1$, as shown in Fig. 1. Light propagates from left to right along the z-axis. Following the notation in Ref [1], we first consider S-polarized evanescent wave for which the electric field is given by,

$$\mathbf{E}_{0S+} = [0,1,0]E_0 \exp(ik_z z + ik_x x - i\omega t), \quad (1)$$

where the wave vector $k_z = +i\sqrt{k_x^2 - k^2}$ and $k = \omega/c < |k_x|$. The electric field of the reflected waves is

$$\mathbf{E}_{0S-} = r[0,1,0]E_0 \exp(-ik_z z + ik_x x - i\omega t), \quad (2)$$

and the electric field of the transmitted waves is,

$$\mathbf{E}_{1S+} = t[0,1,0]E_0 \exp(ik_z z + ik_x x - i\omega t), \quad (3)$$

Both the reflected and transmitted evanescent waves decay away from the interface. Snell's law of refraction has been used to impose that $k_x$ remains the same across the interface. Maxwell's equations relate the electric field to the magnetic field in the media. In air, $i\omega\mu_0 \mathbf{H} = \nabla \times \mathbf{E}$, and in the NIM ($\mu = -1$), $-i\omega\mu_0 \mathbf{H} = \nabla \times \mathbf{E}$. Accordingly,

$$\mathbf{H}_{0S+} = \frac{E_0}{\omega\mu_0}[-k_z, 0, k_x]\exp(ik_z z + ik_x x - i\omega t), \quad (4)$$

$$\mathbf{H}_{0S-} = \frac{rE_0}{\omega\mu_0}[k_z, 0, k_x]\exp(ik_z z + ik_x x - i\omega t), \quad (5)$$

$$\mathbf{H}_{1S+} = \frac{tE_0}{\omega\mu_0}[k_z, 0, -k_x]\exp(ik_z z + ik_x x - i\omega t), \quad (6)$$

Pendry assumed the ordinary boundary conditions, namely, both the tangential components of **E** and **H** are continuous [1]. From Eq.(1) to Eq.(6), this clearly yields $r + 1 = t$ and $r - 1 = t$, which are self-contradictory. It is these self-contradictory equations that resulted in both reflection and transmission coefficients diverging and, following Pendry's treatment, yielded the amplification of evanescent waves going through the NIM slab. Similar self-contradictory equations resulted from using the ordinary boundary conditions for P-polarized evanescent waves the derivation for which follows that for S-polarized light presented here.

## III. IMPACT OF SURFACE CURRENTS AND SURFACE CHARGES

The ordinary boundary conditions used by Pendry and others neglect possible non-zero surface current and charge densities. In the most general case, while the tangential component of the electric field is always continuous, the tangential component of the magnetic field may not be continuous if there is a non-zero surface current density. For example, on the macroscopic scale where distances under consideration are much larger than the wavelength, there are normally both surface charge density and surface current density fluctuations at the metal-dielectric interface. The surface charge density will make the normal component of the electric field discontinuous, and the surface current density will make the tangential component of the magnetic field discontinuous. For S-polarized light input, since the normal components of the electric fields are zero in both media, it can be concluded that the surface charge density $\Sigma$ remains zero. However, the induced surface current density may not be zero for the evanescent wave cases. In fact, assuming the surface current density is $\mathbf{K}$, one has for the tangential components of the magnetic fields,

$$\mathbf{n} \times (\mathbf{H}_1 - \mathbf{H}_0) = \mathbf{K} \tag{7}$$

where $\mathbf{n}$ is the unit vector pointing from air to the NIM. Using Eq.(4) to Eq.(6) for the magnetic fields, one can derive that $K_x = 0$, and

$$K_y = \frac{k_z E_0}{\omega \mu_0}(t - r + 1)\exp(ik_x x - i\omega t) \tag{8}$$

Now since the boundary condition for the tangential component of the electric field is continuous, one still has $r + 1 = t$. Accordingly, one can conclude that for this special case,

$$K_y = \frac{2k_z E_0}{\omega \mu_0}\exp(ik_x x - i\omega t) \tag{9}$$

This surface current density is induced surface current and is zero only when input field is zero.

Following the Green's function method [37]–[39], for $z > 0$ in NIM, we have the vector potential as:

$$\mathbf{A}(\mathbf{r}) = -\mu_0 \int\int \mathbf{K}(\mathbf{r}')\frac{\exp(-ik|\mathbf{r}-\mathbf{r}'|)}{4\pi|\mathbf{r}-\mathbf{r}'|}dS' \tag{10}$$

and the electric field will be:

$$\mathbf{E} = i\omega \mathbf{A} \tag{11}$$

Eq.(10) can be integrated analytically, which yields $A_x = A_z = 0$, and,

$$A_y(x,y,z) = -\frac{k_z E_0}{\omega}\exp(ik_x x - i\omega t)\int_0^\infty J_0(k_x \rho)\frac{\exp\left(-jk\sqrt{\rho^2 + z^2}\right)}{\sqrt{\rho^2 + z^2}}\rho d\rho \tag{12}$$

where $J_0(x)$ is the Bessel function of the 0-th order. For $|k_x| > k$ and $z > 0$, we have [40],

$$A_y(x,y,z) = -\frac{k_z E_0}{\omega\sqrt{k_x^2 - k^2}}\exp(ik_x x - i\omega t)\exp(-\sqrt{k_x^2 - k^2}\, z) \tag{13}$$

Recall $k_z = +i\sqrt{k_x^2 - k^2}$ for evanescent waves

$$A_y(x,y,z) = -\frac{iE_0}{\omega}\exp(ik_x x + ik_z z - i\omega t) \tag{14}$$

Therefore, using Eq.(11), for $z > 0$ in the NIM, $E_x = E_z = 0$, and

$$E_y(x, y, z) = E_0 \exp(ik_x x + ik_z z - i\omega t) \tag{15}$$

Comparing this to Eq.(3), we have $t = 1$. Since the tangential components of the electric field are always continuous, we still have $1 + r = t$. Therefore, given $t = 1$, one has $r = 0$. This analytical solution to the refraction and reflection of evanescent waves from air to the ideal NIM of $n = -1$ is completely self-consistent and satisfactory. The conclusion that $t = 1$ and $r = 0$ also agree with the intuitive analysis where one can argue that since the wave impedance of air and the ideal NIM matches, no reflection of waves should be expected.

A similar analysis can be done for P-polarized evanescent waves. There, we start with magnetic fields,

$$\mathbf{H}_{0P+} = [0,1,0] H_0 \exp(ik_z z + ik_x x - i\omega t), \tag{16}$$

The magnetic field of the reflected wave is,

$$\mathbf{H}_{0P-} = r_H [0,1,0] H_0 \exp(-ik_z z + ik_x x - i\omega t), \tag{17}$$

and the magnetic field of the transmitted wave is,

$$\mathbf{H}_{1P+} = t_H [0,1,0] H_0 \exp(ik_z z + ik_x x - i\omega t), \tag{18}$$

Electric fields can be derived from magnetic fields in air and NIM using $-i\omega\epsilon_0 \mathbf{E} = \nabla \times \mathbf{H}$ and $i\omega\epsilon_0 \mathbf{E} = \nabla \times \mathbf{H}$ respectively. The tangential component of electric field is always continuous. This yields $1 - r_H = -t_H$. The surface current and charge densities in this case can also be determined as follows:

$$K_x = 2H_0 \exp(ik_x x - i\omega t) \tag{19}$$

$$K_y = 0 \tag{20}$$

and

$$\Sigma = \frac{2H_0 k_x}{\omega} \exp(ik_x x - i\omega t) \tag{21}$$

Note these results are also self-consistent and satisfy the continuity equation $\nabla \cdot \mathbf{K} - i\omega\Sigma = 0$.

The vector potential field is still given by Eq.(10) and relates to the magnetic flux and the magnetic field in NIM as,

$$\mathbf{B} = -\mu_0 \mathbf{H} = \nabla \times \mathbf{A} \tag{22}$$

Following similar steps to those above, we can have for $z > 0$ in the NIM, $A_y = A_z = 0$ and,

$$A_x(x, y, z) = -\frac{i\mu_0 H_0}{k_z} \exp(ik_x x + ik_z z - i\omega t) \tag{23}$$

and,

$$H_y(x, y, z) = -H_0 \exp(ik_x x + ik_z z - i\omega t) \tag{24}$$

Comparing to Eq.(18), we have $t_H = -1$ and $r_H = 0$.

A similar analysis can be done at the interface between the NIM and the air when evanescent waves impinge from the NIM side. The conclusion is the same there that all evanescent waves are transmitted and there is no reflection. Accordingly, a slab of the NIM on the macroscopic scale, i.e., with a thickness that is many times that of the wavelength, behaves just like the same thickness of air for any evanescent wave input and there is no amplification effect for the evanescent waves.

DISCUSSION AND CONCLUSION

In summary, we have resolved the problem of evanescent wave refraction and reflection at an interface of air and an ideal NIM of $\epsilon = -1$ and $\mu = -1$. We found the previous treatment by Pendry, which neglects the effect of non-zero induced surface current and/or surface charge densities, is self-contradictory. The self-consistent treatment taking into account the induced surface current and charges yields a result that all evanescent waves are transmitted and there is no reflection. Accordingly, there is no evanescent wave amplification in the NIM slab and the negative index of refraction does not make a perfect lens. This conclusion is not in contradiction with other numerical simulated and experimentally demonstrated super-lens effects since in those the slab thickness is usually a fraction of wavelength, in which case surface plasma polaritons (SPP) may couple efficiently between front and back interface. Accordingly, the demonstrated super-lens effects, or the transport of evanescent wave components, are purely due to the SPP coupling, rather than the evanescent wave amplification by the NIM slab.

REFERENCES:

[1] J. B. Pendry, "Negative Refraction Makes a Perfect Lens," *Phys. Rev. Lett.*, vol. 85, no. 18, pp. 3966–3969, Oct. 2000.
[2] D. R. Smith, "APPLIED PHYSICS: How to Build a Superlens," *Science*, vol. 308, no. 5721, pp. 502–503, 2005.
[3] N. Fang, H. Lee, C. Sun, and X. Zhang, "Sub-Diffraction-Limited Optical Imaging with a Silver Superlens," *Science*, vol. 308, no. 5721, pp. 534–537, 2005.
[4] T. Taubner, D. Korobkin, Y. Urzhumov, G. Shvets, and R. Hillenbrand, "Near-Field Microscopy Through a SiC Superlens," *Science*, vol. 313, no. 5793, p. 1595–, 2006.
[5] I. I. Smolyaninov, Y.-J. Hung, and C. C. Davis, "Magnifying Superlens in the Visible Frequency Range," *Science*, vol. 315, no. 5819, pp. 1699–1701, 2007.
[6] E. Cubukcu, K. Aydin, E. Ozbay, S. Foteinopolou, and C. M. Soukoulis, "Subwavelength Resolution in a Two-Dimensional Photonic-Crystal-Based Superlens," *Phys. Rev. Lett.*, vol. 91, no. 20, p. 207401, Nov. 2003.
[7] H. Lee, Y. Xiong, N. Fang, W. Srituravanich, S. Durant, M. Ambati, C. Sun, and X. Zhang, "Realization of optical superlens imaging below the diffraction limit," *New Journal of Physics*, vol. 7, p. 255, 2005.
[8] D. Melville and R. Blaikie, "Super-resolution imaging through a planar silver layer," *Opt. Express*, vol. 13, no. 6, pp. 2127–2134, 2005.
[9] D. Schurig, J. J. Mock, B. J. Justice, S. A. Cummer, J. B. Pendry, A. F. Starr, and D. R. Smith, "Metamaterial Electromagnetic Cloak at Microwave Frequencies," *Science*, vol. 314, no. 5801, pp. 977–980, 2006.
[10] W. Cai, U. K. Chettiar, A. V. Kildishev, and V. M. Shalaev, *Optical Cloaking with Non-Magnetic Metamaterials*. 2006.
[11] W. Cai, U. K. Chettiar, A. V. Kildishev, V. M. Shalaev, and G. W. Milton, "Nonmagnetic cloak with minimized scattering," *Applied Physics Letters*, vol. 91, no. 11, p. 111105, 2007.
[12] W. Cai, U. K. Chettiar, A. V. Kildishev, and V. M. Shalaev, "Designs for optical cloaking with high-order transformations," *Opt. Express*, vol. 16, no. 8, pp. 5444–5452, 2008.
[13] S. A. Cummer, B.-I. Popa, D. Schurig, D. R. Smith, and J. Pendry, "Full-wave simulations of electromagnetic cloaking structures," *Physical Review E (Statistical, Nonlinear, and Soft Matter Physics)*, vol. 74, no. 3, p. 036621, 2006.
[14] Z. Ruan, M. Yan, C. W. Neff, and M. Qiu, "Ideal Cylindrical Cloak: Perfect but Sensitive to Tiny Perturbations," *Physical Review Letters*, vol. 99, no. 11, p. 113903, 2007.
[15] Z. Liu, H. Lee, Y. Xiong, C. Sun, and X. Zhang, "Far-Field Optical Hyperlens Magnifying Sub-Diffraction-Limited Objects," *Science*, vol. 315, no. 5819, p. 1686–, 2007.
[16] A. Grbic, L. Jiang, and R. Merlin, "Near-Field Plates: Subdiffraction Focusing with Patterned Surfaces," *Science*, vol. 320, no. 5875, pp. 511–513, 2008.

[17] A. Salandrino and N. Engheta, "Far-field subdiffraction optical microscopy using metamaterial crystals: Theory and simulations," *Physical Review B (Condensed Matter and Materials Physics)*, vol. 74, no. 7, p. 075103, 2006.

[18] M. Scalora, G. D'Aguanno, N. Mattiucci, M. J. Bloemer, D. de Ceglia, M. Centini, A. Mandatori, C. Sibilia, N. Akozbek, M. G. Cappeddu, M. Fowler, and J. W. Haus, "Negative refraction and sub-wavelength focusing in the visible range using transparent metallo-dielectric stacks," *Optics Express*, vol. 15, pp. 508–523, Jan. 2007.

[19] B. Wood, J. B. Pendry, and D. P. Tsai, "Directed subwavelength imaging using a layered metal-dielectric system," *Physical Review B (Condensed Matter and Materials Physics)*, vol. 74, no. 11, p. 115116, 2006.

[20] A. Alù and N. Engheta, "Achieving transparency with plasmonic and metamaterial coatings," *Physical Review E (Statistical, Nonlinear, and Soft Matter Physics)*, vol. 72, no. 1, p. 016623, 2005.

[21] A. A. Govyadinov and V. A. Podolskiy, "Metamaterial photonic funnels for subdiffraction light compression and propagation," *Physical Review B (Condensed Matter and Materials Physics)*, vol. 73, no. 15, p. 155108, 2006.

[22] Z. Jacob, I. Smolyaninov, and E. E. Narimanov, "Single photon gun: Radiative decay engineering with metamaterials," 2009, pp. 1–2.

[23] M. A. Noginov, H. Li, Y. A. Barnakov, D. Dryden, G. Nataraj, G. Zhu, C. E. Bonner, M. Mayy, Z. Jacob, and E. E. Narimanov, "Controlling spontaneous emission with metamaterials," *Opt. Lett.*, vol. 35, no. 11, pp. 1863–1865, 2010.

[24] K. Sinchuk, R. Dudley, J. D. Graham, M. Clare, M. Woldeyohannes, J. O. Schenk, R. P. Ingel, W. Yang, and M. A. Fiddy, "Tunable negative group index in metamaterial structures with large formbirefringence," *Opt. Express*, vol. 18, no. 2, pp. 463–472, 2010.

[25] G. W. 't Hooft, "Comment on ar16Negative Refraction Makes a Perfect Lens ar17," *Phys. Rev. Lett.*, vol. 87, no. 24, p. 249701, Nov. 2001.

[26] J. Pendry, "Pendry Replies:," *Phys. Rev. Lett.*, vol. 87, no. 24, p. 249702, Nov. 2001.

[27] J. M. Williams, "Some Problems with Negative Refraction," *Phys. Rev. Lett.*, vol. 87, no. 24, p. 249703, Nov. 2001.

[28] J. Pendry, "Pendry Replies:," *Phys. Rev. Lett.*, vol. 87, no. 24, p. 249704, Nov. 2001.

[29] N. Garcia and M. Nieto-Vesperinas, "Left-Handed Materials Do Not Make a Perfect Lens," *Phys. Rev. Lett.*, vol. 88, no. 20, p. 207403, May 2002.

[30] M. Nieto-Vesperinas and N. Garcia, "Nieto-Vesperinas and Garcia Reply:," *Phys. Rev. Lett.*, vol. 91, no. 9, p. 099702, Aug. 2003.

[31] M. Nieto-Vesperinas, "Problem of image superresolution with a negative-refractive-index slab," *J. Opt. Soc. Am. A*, vol. 21, no. 4, pp. 491–498, Apr. 2004.

[32] D. R. Smith, D. Schurig, M. Rosenbluth, S. Schultz, S. A. Ramakrishna, and J. B. Pendry, "Limitations on subdiffraction imaging with a negative refractive index slab," *Applied Physics Letters*, vol. 82, no. 10, pp. 1506–1508, 2003.

[33] Z. Liu, N. Fang, T.-J. Yen, and X. Zhang, "Rapid growth of evanescent wave by a silver superlens," *Applied Physics Letters*, vol. 83, no. 25, pp. 5184–5186, 2003.

[34] W. Yang, J. O. Schenk, and M. A. Fiddy, "Revisiting Perfect Lens: Negative Refraction Does Not Make a Perfect Lens," *arXiv.org:0807.1768, ıt submitted to J. Opt. Soc. Am. B*, 2009.

[35] W. Yang, J. O. Schenk, and M. A. Fiddy, "Revisiting the perfect lens with loss," in *IEEE SoutheastCon 2010 (SoutheastCon), Proceedings of the*, 2010, pp. 302–304.

[36] Weiguo Yang and Michael A. Fiddy, "Impact of Surface Current on Perfect Lens," in *IEEE SoutheastCon 2013 (SoutheastCon), Proceedings of the*, 2013.

[37] J. D. Jackson, *Classical Electrodynamics*, 2nd ed. John Wiley & Sons, 1975.

[38] R. Harrington, *Field Computation by Moment Methods*. IEEE Press, 1993.

[39] Walton C. Gibson, *The Method of Moments in Electromagnetics*. Taylor & Francis Group, 2008.

[40] I. M. R. I. S. Gradshteyn, *Table of Integrals, Series, and Products*, 7th Ed. Academic Press, Elsevier, 2007.